\documentclass[preprint,12pt]{elsarticle}
 
\usepackage[margin=1in]{geometry} 
\usepackage{amsmath,amsthm,amssymb}
\usepackage[english]{babel}
\usepackage{caption}
\usepackage{subcaption}
\usepackage{tikz-cd}
\usepackage{enumitem}
\setlist[itemize]{nolistsep}
\setlist[enumerate]{label=(\alph*),nolistsep}
\usepackage{hyperref}
\usepackage{bbm}
\usepackage{bm}
\usepackage{physics}
\usepackage[textsize=small]{todonotes}

\journal{Physics Letters A}

\newtheorem{thm}{Theorem}[subsection]

\theoremstyle{definition}

\theoremstyle{remark}

\begin{document}

\begin{frontmatter}

\title{Towards analytical calculation of the magnetic flux measured by magnetometers}

\address[inst1]{Department of Physics, University of Washington, Seattle, WA 98195, USA}
\address[inst2]{Institute for Learning and Brain Sciences, University of Washington, Seattle, WA 98195, USA}
\address[inst3]{Department of Mathematics, University of Pennsylvania, Philadelphia, PA 19104, USA}
            
\author[inst1,inst2]{Wan-Jin Yeo\corref{cor1}}
\ead{wyeo@uw.edu}
\author[inst3]{Yao-Rui Yeo}
\ead{yeya@sas.upenn.edu}
\author[inst1,inst2]{Samu Taulu}
\ead{staulu@uw.edu}

\cortext[cor1]{Corresponding author}

\begin{abstract}
This paper reviews magnetic flux signal calculations through pick-up loops using vector spherical harmonic expansion under the quasi-static approximation, and presents a near-analytical method of evaluating the flux through arbitrary parametrizable pick-up loops for each expansion degree. This is done by simplifying the surface flux integral (2D) into a line integral (1D). For the special case of tangential circular sensors, we present a fully analytical recursion calculation. We then compare commonly-used cubature approximations to our (near-)analytical forms, and show that cubature approximations suffer from increasing errors for higher spatial frequency components. This suggests the need for more accurate evaluations for increasingly sensitive sensors that are being developed, and our (near-)analytical forms themselves are a solution to this problem.
\end{abstract}

\begin{keyword}
Magnetic measurements \sep Magnetoencephalography \sep Pickup loops \sep Surface integral \sep Line integral \sep Vector spherical harmonics
\end{keyword}

\end{frontmatter}

\section{Introduction}
Observations of the magnetic field often consist of measuring the corresponding magnetic flux with instruments such as conducting pick-up loops. In order to apply mathematical models to such measurements, one must calculate the flux as a surface integral of the magnetic field for the given measurement geometry, which is typically conducted by numerical integration. Such calculations are prone to errors that depend on the accuracy of these numerical methods, which is especially critical for spatially complex magnetic field patterns. It is, of course, possible to improve the accuracy by simply increasing the number of integration points, but the evaluation of such integrals may become time-consuming, especially if they have to be performed separately for each possible source configuration of interest. Therefore, it is desirable to develop methodology that provides 1) error estimates of numerical integration for different levels of spatial complexity of the magnetic field, 2) means of simplifying the formalism of integration, 3) or, in certain situations, analytical formulae for the surface integrals. The methods presented in this paper apply generally to the measurement of the flux of quasi-static magnetic fields, but we investigate magnetoencephalography (MEG) instruments in particular. Our methodology also allows the flux integrals to be calculated independently of the source configuration. Hence, the flux values only need to be calculated once for a given sensor geometry, and the contribution of the underlying source distribution can be evaluated by updating certain weight coefficients, the magnetostatic multipole moments, that correspond to the flux values. This is a faster operation than the conventional forward calculation, which requires the so-called lead field basis matrix to be updated for every flux integration point when the source configuration is updated. 

In this paper we focus on MEG, which is a non-invasive technique to measure magnetic fields produced by brain activity \cite{cohen,HHIKL}. These fields  are extremely weak and typically on the order of less than 1000 fT at the measurement sites, hence highly sensitive sensors are required to discernibly pick up these signals. The most commonly employed sensors with sufficient sensitivity are superconducting quantum interference devices (SQUIDs). However, SQUIDs may only be operated at cryogenic temperatures and they are thus contained within dewars typically filled with liquid helium (boiling point of around 4 K) in order to maintain the superconducting state. The necessity of these thick-walled dewars to insulate the head from the liquid helium renders the closest possible distance between SQUID sensors and the scalp to be around 2 cm \cite{opm1}. This distance between the sensors and the scalp is a factor that compromises the detection of high spatial frequency patterns since these weak signals decay fast as a function of distance, and hence sets a limit to the spatial resolution of the fields. The spatial frequency spectrum of the magnetic field detectable by the measuring instrument determines the fundamental resolution of source reconstruction obtained by inverse models.

Recently, advancements have been made in developing optically pumped magnetometer sensors (OPM) \cite{opm1}-\cite{opm11}, which offer a solution to reduce sensor distance from the scalp. OPMs are able to operate at room temperatures, thus they may be placed directly on the scalp. This allows for the detection of signals with higher spatial frequencies, i.e. the OPM sensors may have a higher sensitivity than the SQUID sensors \cite{opm5}. As opposed to flat pick-up loops of SQUID sensors, OPM sensors have a cylindrical sensing volume for which the flux is calculated over, for example a cube \cite{opm1}.

In order for one to exploit the increased spatial information provided by the novel MEG measurement arrangements, it is important to investigate whether the approximation methods of the magnetic flux calculations are accurate enough. This also motivates us to revisit the fundamentals of the flux calculations more generally. The usual method of calculating the magnetic flux signals over pick-up loops is to use numerical cubature integral approximation over the sensor (see \cite{AS}, \cite{opm1}, \cite{DA}). In this paper, we provide an expression to calculate to arbitrary accuracy the magnetic flux contribution from each frequency band corresponding to each harmonic degree $l$ of spherical harmonics. This is facilitated by reducing the surface integral to a line integral. We also show a way to analytically calculate the flux for circular tangential sensors. We show that the current approximation methods suffer from increased errors for signals corresponding to higher spatial frequencies, which correspond to magnetic field components offering potentially better source resolution information than the lower spatial frequencies. Thus, for next-generation OPM sensors that can detect higher spatial frequencies, some of the commonly-used cubature approximations may yield inaccurate flux calculations. The equations provided in this paper that allow for exact evaluations can mitigate this issue for pick-up loops. To our knowledge, this is the first paper that allows for evaluation of MEG flux signals of arbitrary sensor configuration to near-exact accuracy. Our paper also serves as a review on mathematical aspects of  magnetic flux calculations, with emphasis on the vector spherical harmonics (VSH) expansion representation of the magnetic field.

Another issue that may arise when using current cubature approximations to resolve high frequency signals is aliasing. The distances between the discrete sampling points may fall below the spatial Nyquist sampling rate, and hence alias the high frequency signal components. These aliased components will manifest as unexplained noise and decrease signal resolution \cite{alias1, alias2}. Being able to calculate the fluxes analytically without the use of discrete sampling points will avoid this problem entirely.

In Section \ref{sssmethod}, we summarize the multipolar decomposition of MEG signals with VSH, which provides a fundamental way to organize the signal into spatial frequency bands. In Section \ref{generalformalism}, we present a general formula for the pick-up loop flux from each frequency band contribution. The expression is applicable for any sensor geometry as long as we can find a parametrization for it. In Sections \ref{squaresensor} and \ref{circularsensor}, we evaluate the cases for rectangular/square and circular sensors respectively, two of the most common sensor geometries. We offer an additional computationally-efficient analytical recursion to calculate the circular sensor case by exploiting symmetry. In Section \ref{results}, we present the cubature errors for various sensor configurations. In Section \ref{discussion}, we  discuss these errors' implications, and comment on potential future directions. In Section \ref{conclusion} we conclude the work.

\section{VSH expansion and the magnetic flux} \label{sssmethod}

Let us assume quasistatic approximation of Maxwell's equations for electromagnetic fields. In the case of MEG, this is valid since the characteristic length scale of magnetic field spatial variations is much larger than the size of the head \cite{HHIKL}. Moreover, let us assume the region in which sensors are located to be source-free, and let $\mathbf{r} = (R,\theta,\phi)$ be the field coordinates. We can thus write $\nabla \times \mathbf{B}(\mathbf{r}) = 0$, which indicates that $\mathbf{B}$ can be expressed as a gradient of a scalar potential $V$,
\begin{equation}\label{gradV}
    \mathbf{B} (\mathbf{r}) = - \mu_0 \nabla V (\mathbf{r}).
\end{equation}
Together with $\nabla \cdot \mathbf{B} = 0$, we see that $V$ satisfies Laplace's equation $\nabla^2 V = 0$. When solved in spherical coordinates, it gives us an expansion of $V$ in terms of (surface) spherical harmonics $Y_{lm}$,
\begin{equation} \label{Vexp}
    V (\mathbf{r}) = \sum_{l=1}^\infty \sum_{m=-l}^l \alpha_{lm} \frac{Y_{lm} (\theta, \phi)}{R^{l+1}} + \sum_{l=1}^\infty \sum_{m=-l}^l \beta_{lm} R^l Y_{lm} (\theta, \phi),
\end{equation}
where the spherical harmonics are defined by
\begin{equation}\label{sphericalharmonics}
Y_{l,m}(\theta,\phi) = \sqrt{\frac{2l+1}{4\pi}\frac{(l-m)!}{(l+m)!}} e^{im\phi} P_l^m(\cos\theta),
\end{equation}
with $l\in \mathbb{N}$, $-l\leq m\leq l$, and $P_l^m$ are the Legendre polynomials defined by
\begin{equation}
P_l^m(x) = \frac{(-1)^m}{2^l l!}(1-x^2)^{m/2} \frac{d^{m+l}}{dx^{m+l}} (x^2-1)^l.
\end{equation}
Notice that we exclude $l=0$ in \eqref{Vexp} due to the absence of magnetic monopoles. Substituting \eqref{Vexp} into \eqref{gradV}, we have
\begin{align} \label{Bexp}
    \mathbf{B} (\mathbf{r}) &= -\mu_0 \sum_{l=1}^{\infty} \sum_{m=-l}^{l} \alpha_{lm} \frac{\bm{\nu}_{lm} (\theta,\phi) }{R^{l+2}} - \mu_0 \sum_{l=1}^{\infty} \sum_{m=-l}^{l} \beta_{lm} R^{l-1} \bm{\omega}_{lm} (\theta,\phi) \nonumber \\
    &\equiv \mathbf{B}_{in} ( \mathbf{r} ) + \mathbf{B}_{out} ( \mathbf{r} ),
\end{align}
with the coefficients $\alpha_{lm}$ and $\beta_{lm}$ (multipole moments) being \cite{TK}
\begin{align}
    \alpha_{lm} &= \frac{i}{2l+1} \sqrt{\frac{l}{l+1}} \int_{v'} R'^l \mathbf{X}_{lm}^* \cdot \mathbf{J}_{in} \ dv'  \equiv \frac{i}{2l+1} \sqrt{\frac{l}{l+1}} A_{\alpha_{lm}} \label{alphalm}, \\
    \beta_{lm} &= \frac{i}{2l+1} \sqrt{\frac{l+1}{l}} \int_{v'} \frac{\mathbf{X}_{lm}^*}{R'^{l+1}} \cdot \mathbf{J}_{out} \ dv' \equiv \frac{i}{2l+1} \sqrt{\frac{l+1}{l}} A_{\beta_{lm}}. \label{betalm}
\end{align}
In the equations above, $i$ is the imaginary unit, $\bm{\nu}_{lm}$, $\bm{\omega}_{lm}$, $\mathbf{X}_{lm}$ are vector spherical harmonics defined as in \cite{ELH} via the components of $\nabla(Y_{lm}/R^{l+1})$ and $\nabla(R^l Y_{lm})$ (they are shown explicitly in the \ref{appendixVSH} of this paper). $\mathbf{J}_{in}$ and $\mathbf{J}_{out}$ are \emph{total} source current distributions with radial distance smaller or greater than the sensor radial distance respectively, and we have defined $A_{\alpha_{lm}}$ and $A_{\beta_{lm}}$ as the volume integrals over the source space. These multipole moment expressions may be obtained using orthogonality relations of the VSH. The spherical harmonic degrees $l$ correspond to frequency bands; the higher the degree of $l$, the higher the spatial frequency. 

Let primed coordinates $\mathbf{r}' = (R',\theta',\phi')$ refer to source coordinates. As anticipated by our definitions of $\mathbf{B}_{in}$, $\mathbf{B}_{out}$, $\mathbf{J}_{in}$ and $\mathbf{J}_{out}$, the first expansion terms for  \eqref{Vexp} and \eqref{Bexp} correspond to contributions where $R' < R$ (i.e. sources within the head), and the second expansion terms correspond to contributions where $R' > R$ (sources outside the head). This can be seen by how the first expansion term necessarily converges as $1/R^{l+2}$ to zero as $R \rightarrow \infty$, whereas the second converges as $R^{l-1}$ as $R\rightarrow 0$.

The decay behaviour of the first term also shows how having sensors closer to the scalp will detect higher frequencies from contributions within the head. Let $R_c$ be a sensor distance that is closer to the scalp than sensor distance $R_f$; $R_c < R_f$. Then, the relative signal amplitude for each $l$ degree for the further sensor compared to the closer sensor is $((1/R_f)/(1/R_c))^{l+2} = (R_c/R_f)^{l+2}$. Since $R_c / R_f < 1$, this quantity converges to zero as $l \rightarrow \infty$. Especially when $R_c << R_f$, it converges quicker to zero at lower $l$ degrees. This indicates that since high frequency amplitudes are much larger at closer sensor positions compared to further sensor positions, the closer sensor will be more sensitive to the high frequency components.

\subsection{Magnetic flux through planar pick-up loops}
In the case of SQUID MEG systems, planar pick-up loops are utilized for signal detection. The signal $\bm{\Phi}_{surf} $ obtained from a sensor of surface $\mathcal{C}$ oriented with unit normal $\mathbf{n} = (n_x, n_y, n_z )$ is the magnetic flux across $\mathcal{C}$, 
\begin{equation}\label{originalflux}
     \bm{\Phi}_{surf} = \int_\mathcal{C} \mathbf{B} (\mathbf{r}) \cdot \mathbf{n} dS,
\end{equation}
where $dS$ refers to an infinitesimal surface element. Due to their equivalence, we will use ``signal'' and ``flux'' interchangeably in this paper. Note that we have also denoted surface fluxes for SQUID pick-up loops as $\bm{\Phi}_{surf}$; for volumetric fluxes of OPM sensors which we will consider later, we denote them as $\bm{\Phi}_{vol}$.

From \eqref{Bexp}, we see that the signal $\bm{\Phi}_{surf}$ can similarly be separated into two components $\bm{\Phi}_{in}$ and $\bm{\Phi}_{out}$ corresponding to contributions from sources within and outside the head,
\begin{align} \label{phiinout}
    \bm{\Phi}_{surf} &= \int_\mathcal{C} \mathbf{B}_{in} (\mathbf{r})  \cdot \mathbf{n} dS + \int_\mathcal{C} \mathbf{B}_{out} (\mathbf{r})  \cdot \mathbf{n} dS \nonumber \\
    &\equiv \bm{\Phi}_{in} + \bm{\Phi}_{out} \\
    &= -\mu_0 \sum_{l=1}^\infty \sum_{m=-l}^l \alpha_{lm} \int_\mathcal{C} \frac{\bm{\nu}_{lm} (\theta,\phi)}{R^{l+2}} \cdot \mathbf{n} dS -\mu_0 \sum_{l=1}^\infty \sum_{m=-l}^l \beta_{lm} \int_\mathcal{C} R^{l-1} \bm{\omega}_{lm} (\theta,\phi) \cdot \mathbf{n} dS. \nonumber
\end{align}
This can be written more compactly in matrix form as
\begin{equation}
    \bm{\Phi}_{surf} = \begin{bmatrix}
    \mathbf{S}_{in} \ \mathbf{S}_{out}
    \end{bmatrix} \begin{bmatrix}
    \mathbf{a} \\
    \mathbf{b}
    \end{bmatrix},
\end{equation}
where
\begin{align}
    \mathbf{S}_{in}  &= \int_\mathcal{C} \left[\frac{\bm{\nu}_{1,-1}}{R^3} \ , \ \frac{\bm{\nu}_{1,0}}{R^3} \ , \ \frac{\bm{\nu}_{1,1}}{R^3} \ , \ \frac{\bm{\nu}_{2,-2}}{R^4} \ , \ \dots \right] \cdot \mathbf{n} dS, \label{Sin} \\
     \mathbf{S}_{out}  &= \int_\mathcal{C} \left[\bm{\omega}_{1,-1} \ , \ \bm{\omega}_{1,0} \ , \ \bm{\omega}_{1,1} \ , \ R\bm{\omega}_{2,-2} \ , \ \dots \right] \cdot \mathbf{n} dS, \label{Sout} \\
    \mathbf{a} &= -\mu_0 \left[\alpha_{1,-1} \ , \ \alpha_{1,0} \ , \ \alpha_{1,1} \ , \ \alpha_{2,-2} \ , \ \dots \right]^T, \\
    \mathbf{b} &= -\mu_0 \left[\beta_{1,-1} \ , \ \beta_{1,0} \ , \ \beta_{1,1} \ , \ \beta_{2,-2} \ , \ \dots \right]^T. \label{multipoleb} 
\end{align}
This matrix representation is the foundation of the  Signal Space Separation (SSS) methodology as first presented in \cite{TK}. In this form, we see that $\mathbf{S}_{in}$ and $\mathbf{S}_{out}$ contain the bases that span the space of signal contributions due to sources located inside and outside the head, respectively. The basis matrices depend on the sensor geometry only, and information about the source distribution is contained in the multipole moments, which are elements of the vectors $\mathbf{a}$ and $\mathbf{b}$. Thus, the basis matrices can be constructed solely with knowledge about the sensor geometry. They contain information about the sensitivity of the sensors to spatial frequencies, as the column vectors are specified according to the ordered $l$ components by virtue of \eqref{Bexp}. Therefore, the elements of this matrix are what we aim to accurately calculate.

Typically, we are only interested in signals due to activity within the head, so from here on, we only consider $\mathbf{S}_{in}$ which specify $\bm{\Phi}_{in}$. The formalism that is to follow is analogous for $\mathbf{S}_{out}$ as well, so it suffices to exclusively consider $\mathbf{S}_{in}$; the only difference is in the coefficients. For a realistic implementation, we must truncate the series at an appropriate degree of $l = L$ so that the matrices are finite-dimensional and preferably over-determined. This truncation degree is chosen so that the signal is represented sufficiently accurately by the bases; it has been determined that for $\mathbf{S}_{in}$, a truncation degree of $L = 8$ is sufficient for SQUID-based MEG systems \cite{TK}.

In our description above, we have only considered one sensor, so the $\mathbf{S}_{in}$ and $\mathbf{S}_{out}$ basis matrices are $1 \times (L^2 - 1)$-dimensional, and the multipole moment vectors are $(L^2-1) \times T$-dimensional, where $T$ is the number of temporal sampling points. For more realistic applications with an array of $N$ sensors, the full basis matrices will comprise $N$ such rows corresponding to each sensor, i.e. the basis matrices will be $N \times (L^2 - 1)$-dimensional, and the multipole moment vectors will be $(L^2-1) \times T$-dimensional.

\subsubsection{The signal basis elements}
Denote each entry of $\mathbf{S}_{in}$ as $v_{lm}$. Converting $\bm{\nu}_{lm}$ from spherical coordinates to Cartesian coordinates, then carrying out the dot product with sensor orientation $\mathbf{n}$, we have
\begin{equation}\label{fullvlm}
    v_{lm} = \int_\mathcal{C} \frac{1}{R^{l+2}} \left[-(l+1) Y_{lm} c_R (\theta,\phi) + \frac{\partial Y_{lm}}{\partial \theta} c_\theta (\theta,\phi) + \frac{imY_{lm}}{ \sin \theta} c_\phi (\theta,\phi) \right] dS,
\end{equation}
where 
\begin{align}
    c_R (\theta,\phi) &= n_x \sin \theta \cos \phi + n_y \sin \theta \sin \phi + n_z \cos \theta, \label{vlmcx} \\
    c_\theta (\theta,\phi) &= n_x \cos \theta \cos \phi + n_y \cos \theta \sin \phi - n_z \sin \theta, \\
    c_\phi (\theta,\phi) &= - n_x \sin \theta + n_y \cos \phi. \label{vlmcz}
\end{align}
The accurate evaluation of these signal basis matrix elements is beneficial for many applications of the VSH expansion. For instance, it allows for more accurate forward flux calculation  $\bm{\Phi}_{in} = \mathbf{S}_{in} \mathbf{a}$ (up to $L$, and assuming we have knowledge about source and thus can calculate $\mathbf{a}$). It also provides effective filtering of external interference in the SSS applications. As mentioned in the introduction, the current method to calculate \eqref{fullvlm} is via cubature formulas, which will cause increasing basis error for higher frequency $l$ components as we will see in Section \ref{results}. Since next-generation sensors will be able to detect higher frequencies, there is an increased importance to evaluate $v_{lm}$ in more accurate ways.

\subsection{Volumetric magnetic flux through cylindrical sensing volumes}
As opposed to a flux measurement across a pick-up loop like for SQUID sensors, the signal measured by OPM sensors is the volumetric flux of $\mathbf{B}$ across the sensing volume $\mathcal{V}$, projected along some sensing direction $\tilde{\mathbf{n}}$. The sensing direction can be modulated for OPM sensors, and is independent of sensor coordinates. Hence, instead of a surface integral in \eqref{originalflux}, we have a volume integral
\begin{align}
     \bm{\Phi}_{vol} &= \int_\mathcal{V} \mathbf{B} (\mathbf{r}) \cdot \tilde{\mathbf{n}} dV.
\end{align}
We may express $\tilde{\mathbf{n}}$ as a rotation of $\mathbf{n}$, i.e. $\tilde{\mathbf{n}} = \mathcal{R} \mathbf{n}$, by constructing $\mathcal{R}$ using Rodrigues' rotation formula as stated in \ref{appendixrotation}. The rotation axis is $\mathbf{k} = \mathbf{n} \times \tilde{\mathbf{n}}$, and the angle of rotation is the angle $\theta'$ between $\mathbf{n}$ and $\tilde{\mathbf{n}}$. So the signal is equivalently
\begin{equation}
    \bm{\Phi}_{vol} = \int_\mathcal{V} \mathbf{B} (\mathbf{r}) \cdot \mathcal{R} \mathbf{n} dV.
\end{equation}
We may discretize this volume integral as a sum over the contributions of $k$ cross-sectional ``slices'' of $\mathcal{V}$. Then, we will have a sum of $k$ surface integrals over the cross-sectional areas $\mathcal{C}_i$ multiplied by their thicknesses $\Delta w_i$, 
\begin{equation}
     \bm{\Phi}_{vol} \approx \sum_{i=1}^k \Delta w_i \int_{\mathcal{C}_i} \mathbf{B} (\mathbf{r}) \cdot \mathcal{R} \mathbf{n} dS_i.
\end{equation}
In this case, all formalism in this paper holds up to a constant, since entries of $\mathcal{R}$ are all constant. So we may proceed by just considering the case for SQUID sensors. Note that for the above construction of the matrix form as in \eqref{Sin}-\eqref{multipoleb}, we expect to require a higher truncation value $L$ due to higher signal spatial resolution for the OPM sensors.

In Section \ref{discussion}, we propose a starting point for an investigation into analytical evaluations of the volume integral, using cylindrical harmonics.

\section{Line integral formula for the magnetic flux of sensors with arbitrary geometry} \label{generalformalism}
The magnetic vector potential $\mathbf{A}$ is defined as 
\begin{equation}
    \nabla \times \mathbf{A} = \mathbf{B}.
\end{equation}
If we can find an $\mathbf{A}$, then we can apply Stoke's theorem to convert  the flux integral \eqref{originalflux} to a line integral over the boundary $\partial \mathcal{C}$ of $\mathcal{C}$,
\begin{equation} \label{lineintflux}
    \bm{\Phi}_{surf} = \oint_{\partial \mathcal{C}} \mathbf{A} \cdot d \mathbf{l} = \int_{t_1}^{t_2} \mathbf{A} (\mathbf{r}(t)) \cdot \mathbf{r}'(t) dt,
\end{equation}
where $d\mathbf{l}$ points along $\partial \mathcal{C}$ according to orientation $\mathbf{n}$, and $\mathbf{r}(t) = (r_x(t), r_y(t), r_z(t))$ is a parametrization of $\partial \mathcal{C}$ in $t \in [t_1,t_2]$. From \cite{ELH}, we know that 
\begin{equation}
    \nabla \times [f(R) \mathbf{X}_{lm}] = i \sqrt{\frac{l}{2l+1}} \left(\frac{df}{dR} - \frac{l}{R} f \right) \mathbf{V}_{lm} + i \sqrt{\frac{l+1}{2l+1}} \left(\frac{df}{dR}+\frac{l+1}{R} f \right) \mathbf{W}_{lm}.
\end{equation}
By comparing coefficients with the $\mathbf{B}$ field expansion \eqref{Bexp} and substituting in \eqref{alphalm} and \eqref{betalm}, we can solve for $f(R)$ and thus find $\mathbf{A}$,
\begin{align}
    &\begin{cases}
    \displaystyle - \mu_0 A_{\alpha_{lm}} \frac{1}{R^{l+2}} = \frac{d f}{dR} - \frac{l}{R} f \\[10pt]
    \displaystyle - \mu_0 A_{\beta_{lm}} R^{l-1} = \frac{d f}{dR} + \frac{l+1}{R} f
    \end{cases} \\
    f &= \frac{\mu_0}{2l+1}\left( \frac{A_{\alpha_{lm}}}{R^{l+1}} - A_{\beta_{lm}} R^l \right) \\
    \mathbf{A} (\mathbf{r}) &= \sum_{l=1}^\infty \sum_{m=-l}^l \frac{\mu_0}{2l+1}\left( \frac{A_{\alpha_{lm}}}{R^{l+1}} - A_{\beta_{lm}} R^l \right) \mathbf{X}_{lm} (\theta,\phi). \label{vectorpotential}
\end{align}
For sources in the head only, $A_{\beta_{lm}} = 0$. As mentioned in Section \ref{sssmethod}, the case for sources outside the head can be done in an analogous way. We will hence drop the ``$in$'' and ``$out$'' subscripts from now on, since we will only refer to the former. In Cartesian coordinates, $\mathbf{x}_{lm} = -\sqrt{l(l+1)} \mathbf{X}_{lm}$ is
\begin{align}\label{xlmcartesian}
    \mathbf{x}_{lm} &= \left(\frac{m Y_{lm}}{ \tan \theta} \cos \phi - i \frac{\partial Y_{lm}}{\partial \theta} \sin \phi\right) \mathbf{e}_x + \left( \frac{m Y_{lm}}{\tan \theta} \sin \phi + i \frac{\partial Y_{lm}}{\partial \theta} \cos \phi \right) \mathbf{e}_y -m Y_{lm} \mathbf{e}_z \nonumber \\
    &\equiv x_1 (\theta,\phi) \mathbf{e}_x +  x_2 (\theta,\phi) \mathbf{e}_y +  x_3 (\theta,\phi) \mathbf{e}_z.
\end{align}
Equations \eqref{alphalm}, \eqref{lineintflux}, \eqref{vectorpotential} and \eqref{xlmcartesian} indicate that the flux contribution of a frequency component corresponding to degree $l$ is (if we place the ``$surf$'' label as a superscript now)
\begin{equation}\label{parametricphilm}
    \bm{\Phi}^{surf}_{lm} = (-\mu_0 \alpha_{lm}) \left[\frac{1}{li} \int_{t_1}^{t_2} \frac{x_1(\theta,\phi) r'_x + x_2(\theta,\phi) r'_y + x_3(\theta,\phi) r'_z }{R^{l+1}} dt \right],
\end{equation}
where 
\begin{equation}
	R(t) =\sqrt{r_x^2+r_y^2+r_z^2},\qquad \theta(t) =\arccos{\frac{r_z}{\sqrt{r_x^2+r_y^2+r_z^2}}},\qquad \phi(t)=\arctan\frac{r_y}{r_x}.
\end{equation}
Note that the line integral refers to the sensor geometry only and is thus independent of the source configuration. Source-specific calculations refer to the multipole moments $\alpha_{lm}$ only, which can be modeled without reference to the sensors. Also, note that in the evaluation of the coordinate conversions, the $\arctan$ function must be defined appropriately to match polar coordinates. In particular, if $r_x<0$, we need to add or subtract $\pi$ if $r_y>0$ or $r_y<0$ respectively.

It is clear from \eqref{parametricphilm} that since the expression contained within the parenthesis are elements of $\mathbf{a}$, the expression within the square brackets is an equivalent expression of \eqref{fullvlm},
\begin{equation} \label{parametricvlm}
    v_{lm} = \frac{1}{li} \int_{t_1}^{t_2} \frac{x_1(\theta,\phi) r'_x + x_2(\theta,\phi) r'_y + x_3(\theta,\phi) r'_z }{R^{l+1}} dt.
\end{equation}
Hence, if we are able to find a parametrization $\mathbf{r}(t)$ of the boundary an arbitrarily-shaped sensor, we are in practice able to find the exact magnetic flux across it by evaluating the line integral. In this form, we have reduced the number of integration parameters from two (a surface integral) in \eqref{fullvlm} to one (a line integral), which helps computing software to evaluate it more easily, up to possible software round-off errors. In the case of Matlab, we may define a tolerance when using the numerical line integral function, hence this integral may be evaluated up to arbitrary accuracy. One should note, however, that the integral evaluation for complicated parametrizations may still be computationally demanding. For a fixed sensor configuration, we will only need to calculate the basis matrix once; once the basis is constructed, further calculations only require updating the multipole moments which is computationally fast.

\section{Line integral formula for rectangular/square sensors} \label{squaresensor}
For a rectangular sensor, the line integral in \eqref{lineintflux} is the sum over the integrals over the 4 straight edges.

Let the center of the rectangle be $\mathbf{r}_C$, and let capital $X$, $Y$ and $Z$ denote the local sensor Cartesian coordinates. Then, the unit basis vectors (i.e. the sensor $X$, $Y$ orientations) that span the plane containing the sensor area can be written in terms of the global coordinates as $\mathbf{n}_X = (n_{Xx},n_{Xy},n_{Xz})$ and $\mathbf{n}_Y = (n_{Yx},n_{Yy},n_{Yz})$. The unit normal $\mathbf{n}_Z = (n_{Zx},n_{Zy},n_{Zz})$ is equivalent to $\mathbf{n}$ as defined before. Also, let the half-width of the pair of edges parallel to $\mathbf{n}_X$ be $d_X$, and the other pair parallel to $\mathbf{n}_Y$ be $d_Y$. We find the coordinates of the 4 corner points as follows: 
\begin{equation}
    \mathbf{r}_0 = \mathbf{r}_C \pm d_X \mathbf{n}_X \pm d_Y \mathbf{n}_Y.
\end{equation}
These are the starting points of each of the 4 line integrals. The parametrization for one of the edges is thus 
\begin{align}
    \mathbf{r}(t) &= \mathbf{r}_0 + \mathbf{n}_\tau  t = (r_x(t), r_y(t), r_z(t)) , \qquad t \in [0,2d_\tau], \\
    \mathbf{r}'(t) &= \mathbf{n}_\tau,
\end{align}
where $\tau = X$ or $Y$, chosen appropriately depending on which edge we are integrating over. From \eqref{parametricvlm}, the contribution to $v_{lm}$ from the edge is thus
\begin{align} \label{sqvlm}
    v_{lm}^{edge} = \frac{1}{li} \int_0^{2d_\tau} \frac{ x_1 (\theta,\phi) n_{\tau x} + x_2 (\theta,\phi) n_{\tau y} + x_3 (\theta,\phi) n_{\tau z} }{R^{l+1}} dt.
\end{align}
The sum of 4 such integrals gives $v_{lm}$ for rectangular sensors. For square sensors, $d_X=d_Y$.

\section{Formulas for circular sensors} \label{circularsensor}

\subsection{Applying the line integral}

Consider a circular sensor of radius $d$ and center $\mathbf{r}_C$ lying on a plane spanned by orthogonal unit vectors $\mathbf{v}_1$ and $\mathbf{v}_2$. Let the unit normal of the plane be $\mathbf{n}$. The parametrization for the circle is given by 
\begin{align}
    \mathbf{r}(t) &= \mathbf{r}_C + d( \mathbf{v}_1  \cos t + \mathbf{v}_2 \sin t), \qquad t \in [0,2 \pi], \\
    \mathbf{r}'(t) &=  d (- \mathbf{v}_1  \sin t + \mathbf{v}_2 \cos t).
\end{align}
Note that the sine and cosine functions might be interchanged to ensure the correct orientation $\mathbf{n}$. We thus have 
\begin{equation}
    v_{lm} = \frac{1}{li} \int_{0}^{2\pi} \frac{x_1(\theta,\phi) r'_x + x_2(\theta,\phi) r'_y + x_3(\theta,\phi) r'_z }{R^{l+1}} dt.
\end{equation}
Due to the slightly more complicated form of $\mathbf{r}'(t)$, the computational time for this integral may be long.

This can be mitigated by noting that we may actually express $v_{lm}$ as a computationally-efficient analytic recursion for tangential circular sensors. This is achieved by exploiting the symmetry of the circular sensor and various recursions of $Y_{lm}$ and $P_l^m$ using \eqref{fullvlm} itself; the line integral form \eqref{parametricvlm} of $v_{lm}$ does not necessarily simplify the surface integral in this case.

\subsection{Applying the surface integral} \label{circularsensoryr}

Consider a circular sensor of radius $d$ with its center lying on the $z$-axis, $\mathbf{r}_C = (0,0,r_C)$. Assume the sensor is tangential relative to the origin, i.e. $\mathbf{r}_C \cdot \mathbf{n} = r_C$. In this case, $n_x = n_y = 0$ and $n_z = 1$ in equations \eqref{vlmcx}-\eqref{vlmcz}. In Cartesian coordinates, \eqref{fullvlm} simplifies to 
\begin{equation} \label{keyeqn}
	v_{l,m} = \int_{\mathcal{C}} \frac{1}{R^{l+2}} \left[-(l+1)Y_{l,m}(\theta,\phi)\cos\theta - \frac{\partial Y_{l,m}(\theta,\phi)}{\partial\theta}\sin\theta \right] dxdy,
\end{equation}
where
\begin{equation} \label{cartosph}
	R=\sqrt{x^2+y^2+r_C^2},\qquad \theta=\arccos{\frac{r_C}{\sqrt{x^2+y^2+r_C^2}}},\qquad \phi=\arctan\frac{y}{x}.
\end{equation}
Again, we must ensure that the $\arctan$ function is defined appropriately.

\subsubsection{Rewriting the integral}

We now aim to give a purely recursive formula for Equation \eqref{keyeqn}, allowing us to quickly compute $v_{lm}$ without any approximations using numerical integration.

By symmetry of the $e^{im\phi}$ term in $Y_{lm}$, one observes that $v_{lm}=0$ whenever $m\neq 0$. Note that this is not true in general for any sensor geometry; \ref{appendixmzero} shows that this does not hold for square sensors. We now concentrate on the case $m = 0$. For non-negative integers $a,b,u,v$, define
\begin{equation}
	\gamma_{a,b,u,v} := 
		\int_0^d \frac{\zeta^a}{(\zeta^2+r_C^2)^{b/2}} P_v^{0,u}(\frac{r_C}{\sqrt{\zeta^2+r_C^2}}) \, d\zeta,
\end{equation}
where $P_v^{0,u}(x)$ is the $u^{th}$ derivative of $P_v^0(x)$. Note that $\gamma_{a,b,u,v}$ equals $0$ whenever $u>v$.

By substituting $x=\zeta\cos\psi$ and $y=\zeta\sin\psi$, and using the identity 
\begin{equation}
    \frac{\partial Y_{l,0}(\theta,\phi)}{\partial \theta}
    =
    \sqrt{l(l+1)}e^{-i\phi}Y_{l,1}(\theta,\phi),
\end{equation}
equation \eqref{keyeqn} for $m=0$ becomes
\begin{align} \label{keyeqnalt}
	v_{l,0} 
    &=
        \sqrt{\frac{2l+1}{4\pi}} \int_0^{2\pi} \int_0^d \frac{1}{(\zeta^2+r_C^2)^{(l+2)/2}} \bigg[-(l+1)P_l^{0}(\frac{r_C}{\sqrt{\zeta^2+r_C^2}})\frac{r_C}{\sqrt{\zeta^2+r_C^2}} 
        \nonumber \\	
    &\hspace{20em}
    -  P_l^{1}(\frac{r_C}{\sqrt{\zeta^2+r_C^2}})\frac{\zeta}{\sqrt{\zeta^2+r_C^2}} \bigg] \zeta \, d\zeta d\psi
        \nonumber \\	
    &= 2\pi\sqrt{\frac{2l+1}{4\pi}}
    	\left(
    		-(l+1)r_C \gamma_{1,l+3,0,l}
    		+ \gamma_{3,l+4,1,l}
    	\right) .
\end{align}
If $l<1$, then the second term vanishes.

We now recognize that computing $v_{l,0}$ amounts to providing an algorithm to calculate $\gamma_{a,b,u,v}$. Let us recall two well-known recurrences for Legendre polynomials.

\begin{thm}[Bonnet's Formulas]
The Legendre polynomials $P_l^0(x)$ satisfy the recurrences
	\begin{align}
	P_{l+1}^0(x) &= \frac{2l+1}{l+1} x P_{l}^0(x) - \frac{l}{l+1} P_{l-1}^0(x), \\
	\frac{d P_{l+1}^0(x)}{dx} &= (l+1)P_l^0(x)+ x \frac{d P_l^0(x)}{dx}.
	\end{align} 
\end{thm}

By the definition of $\gamma_{a,b,u,v}$, we get the relations
	\begin{align}
	\gamma_{a,b,1,l+1} &= (l+1) \gamma_{a,b,0,l} + r_C \gamma_{a,b+1,1,l}, \label{keyrelation1} \\
	\gamma_{a,b,0,l+1} &= \frac{2l+1}{l+1} r_C \gamma_{a,b+1,0,l} - \frac{l}{l+1} \gamma_{a,b,0,l-1}. \label{keyrelation2}
	\end{align}
If we repeatedly apply equation \eqref{keyrelation1} to itself (by substituting into the second term on the right hand side),
    \begin{align}
    \gamma_{a,b,1,l+1} &= (l+1) \gamma_{a,b,0,l} + r_C \gamma_{a,b+1,1,l} \nonumber   \\
    &= (l+1) \gamma_{a,b,0,l}
    + r_C l \gamma_{a,b+1,0,l-1}
    + r_C^2 \gamma_{a,b+2,1,l-1} \nonumber \\
    &= \dotsm \nonumber \\
    &=
    (l+1) \gamma_{a,b,0,l}
    + r_C l \gamma_{a,b+1,0,l-1}
    + r_C^2 (l-1) \gamma_{a,b+2,0,l-2}
    +\dotsm
    +
    r_C^{l} \gamma_{a,b+l,0,0}. \label{keyrelation1a}
    \end{align}
Notice the last term is absent as $\gamma_{a,b+l,1,0}=0$.

Equations \eqref{keyrelation1} and \eqref{keyrelation2} reduce our problem to computing the integrals
	\begin{equation} \label{keycomputation1}
	\gamma_{a,b,0,0}=
	\int_0^d \frac{\zeta^a}{(\zeta^2+r_C^2)^{b/2}} \, d\zeta
	\end{equation}
and
	\begin{align}  \label{keycomputation2}
	\gamma_{a,b,0,1}
	&= \int_0^d \frac{\zeta^a r_C}{(\zeta^2+r_C^2)^{(b+1)/2}} \, d\alpha  \zeta \\
	&= r_C \gamma_{a,b+1,0,0}.
	\end{align}

\subsubsection{A recurrence relation}

The one-dimensional integral $\gamma_{a,b,0,0}$ \eqref{keycomputation1} can be easily evaluated, but we can also give a recipe to recursively compute it more efficiently.

In light of the discussion above, let us write $\gamma_{a,b}:=\gamma_{a,b,0,0}$. We also restrict ourselves to the case $b\geq a+2$ by equation \eqref{keyeqnalt}. Integrating by parts, one observes that
	\begin{equation} \label{rr1}
	\gamma_{a+2,b+2}=\frac{(a+1)}{b} \gamma_{a,b} - \frac{d^{a+1}}{b(d^2+r_C^2)^{b/2}}.
	\end{equation}
So, it suffices to compute $\gamma_{1,b}$ and $\gamma_{0,b}$. The former component has the form
	\begin{equation} \label{rr2}
	\gamma_{1,b} = -\frac{1}{b-2}\left(\frac{1}{(d^2+r_C^2)^{(b-2)/2}}-\frac{1}{r_C^{b-2}}\right).
	\end{equation}
The substitution $\zeta=r_C \tan\omega$ into \eqref{keycomputation1} gives us 
	\begin{equation} \label{rr3}
	\gamma_{0,b} = \frac{1}{b-2}\left(\frac{b-3}{r_C^2}\gamma_{0,b-2} + \frac{d}{r_C^{2}(d^2+r_C^2)^{(b-2)/2}}\right), \qquad b\geq 4,
	\end{equation}
with base cases
\begin{equation}\label{basecases}
    \gamma_{0,2} = \frac{1}{r}\arctan\left(\frac{d}{r_C}\right) , \qquad \gamma_{0,3} = \frac{d}{r^2\sqrt{d^2+r_C^2}}. 
\end{equation}

Equations \eqref{keyeqnalt}, \eqref{keyrelation2}, \eqref{keyrelation1a}, \eqref{keycomputation2}-\eqref{basecases}, when evaluated in reverse order, provide an efficient way to compute $v_{lm}$ analytically as desired.

\subsubsection{Generalizing to arbitrary tangential sensors}
To generalize the above to sensors not aligned on the $z$ axis, we may passively rotate the $\mathbf{B}$ field so that the $z$ axis now aligns with the center of the new sensor. One way to do so is described in \ref{appendixrotation}. This alters the coefficients and coordinates of $v_{lm}$, but eventually we see that it retains its general form of \eqref{keyeqn} apart from different coefficients. See \ref{appendixrotationcircle} for more details. As such, the formalism to establish a recursion still holds. 

We may employ a similar idea to calculate the line integral flux \eqref{sqvlm} for tangential square sensors. We may passively rotate the magnetic vector potential $\mathbf{A}$ so that the $z$-axis aligns with a sensor of interest, then evaluate the line integral \eqref{sqvlm}. In this case, we will only need to evaluate line integrals with $m$ a multiple of $4$ (including $m=0$); this is shown in \ref{appendixmzero}. This cuts down the number of integrals we need to compute by a quarter, though at the expense of constructing the rotation matrix $\mathcal{R}$ and re-defining the integrand coefficients. For both square and circular sensors, evaluating only the nonzero $v_{lm}$ terms may also improve accuracy of flux calculation, since possible computer software round-off errors from calculating the zero terms will be eliminated. Again, this is done at the expense of introducing round-off errors from the construction of $\mathcal{R}$.

\section{Simulation results} \label{results}

We consider the simplest case of tangential sensors oriented along the $z$-axis, since it is sufficient to illustrate our purposes. For square sensors, we used the line integral \eqref{sqvlm} to obtain near-exact calculations. We determined that default Matlab tolerances gave results that were consistent with decreased tolerances, hence was sufficient for a near-exact evaluation. For circular sensors, we used the recursion in Section \ref{circularsensoryr} for analytical calculations. For cubature approximations, we followed the convention provided in Section 25.4 of \cite{AS}. Namely, we used 4- and 9-point cubatures for square sensors, and 4-, 7- and 21-point cubatures for circular sensors. In addition, for both sensor geometries, we considered the commonly-used point-like sensor, i.e. a 1-point cubature, with the sampling point located at the sensor center. Unless explicitly stated, the distance of the sensor from the origin is $r_C = 9$ cm, which corresponds to the average size of an adult head, and the half-width/radius of the sensor is $d = 1$ cm, approximately the half-width of the square pick-up loops of the Elekta Neuromag TRIUX system (Megin, Helsinki, Finland) \cite{opm5}.

\subsection{Errors for each \texorpdfstring{$l$}{} degree} \label{results1}

\begin{figure}[ht]
    \centering
    \begin{subfigure}[t]{0.49\textwidth}
        \centering
         \includegraphics[width=\textwidth]{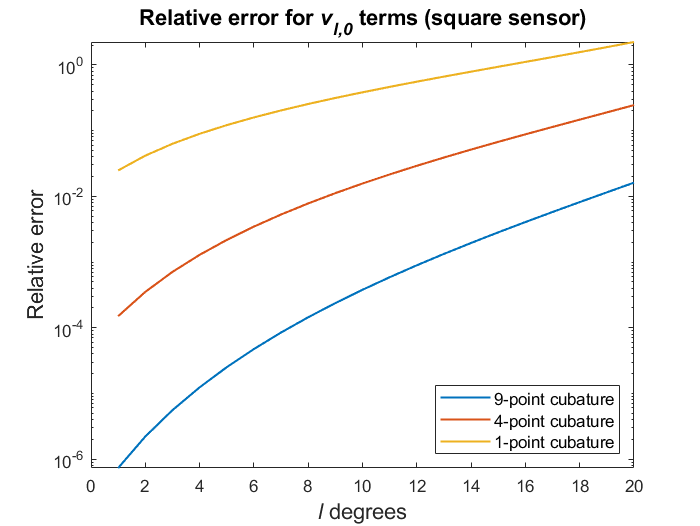}
         \caption{}
    \end{subfigure}
    \begin{subfigure}[t]{0.49\textwidth}
        \centering             
        \includegraphics[width=\textwidth]{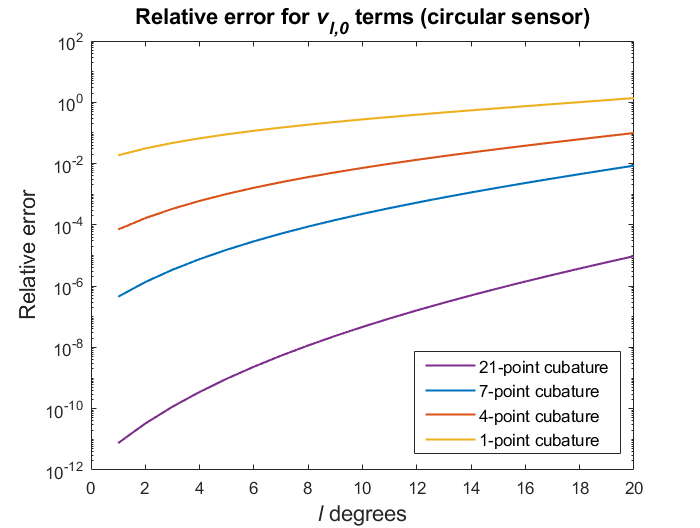}
        \caption{}
    \end{subfigure}
    \caption{Relative error plots for tangential (a) square sensors and (b) circular sensors along the $z$-axis for different $l$ degrees. The errors increase as $l$ increases, which means that flux calculations for higher frequencies become more inaccurate. Fewer sampling points result in larger errors, as expected.}
    \label{fig: figm0err}
\end{figure}

We consider $v_{l,0}$ terms, since they are the only nonzero terms for the circular sensors. Figure \ref{fig: figm0err} shows the relative error of $v_{l,0}$ numerically approximated using the various cubatures with respect to our exact evaluations. As expected, lower number of sampling points yield the greatest errors, and errors increase as $l$ increases, indicating a higher sensitivity of higher spatial frequencies to numerical approximation errors. 

For point-like approximations, there are significant errors of around 16\% and 11\% for square and circular sensors respectively, at $l=6$. This is below the usual truncation of $l=8$, and hence indicates that even for current SQUID sensors, using point-like sensor approximations may introduce significant errors in calculations. The errors for the other cubatures had relatively small errors up to $l=8$, but up to $l=20$ as shown in the plots, only 9-point cubature for square sensors, and 7- and 21-point cubatures for circular sensors had errors $<2\%$.

\subsection{Decreased sensor distance}

\begin{figure}[ht]
    \centering
    \begin{subfigure}[t]{0.49\textwidth}
        \centering
         \includegraphics[width=\textwidth]{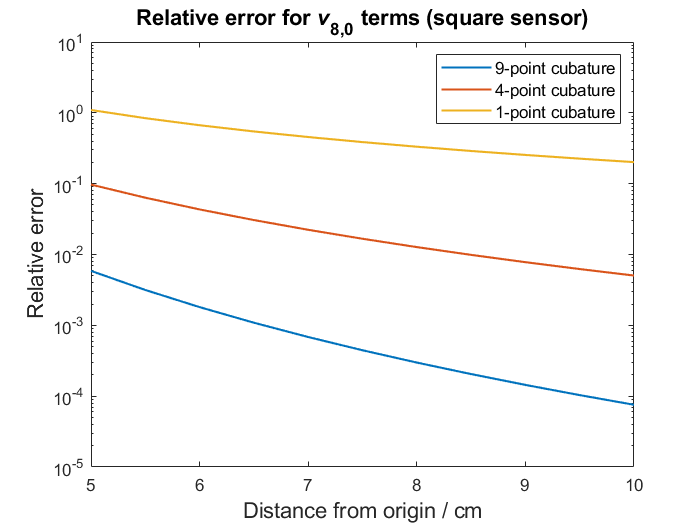}
         \caption{}
    \end{subfigure}
    \begin{subfigure}[t]{0.49\textwidth}
        \centering             
        \includegraphics[width=\textwidth]{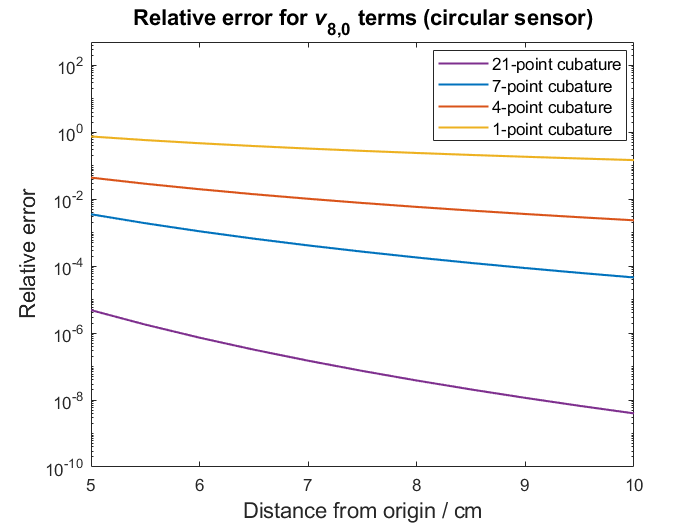}
        \caption{}
    \end{subfigure}
    \caption{Relative error plots for tangential (a) square sensors and (b) circular sensors along the $z$-axis for various sensor distances from the origin. As expected, errors increase as the sensor is placed closer to the origin, since higher frequency components correspondingly increase in amplitude.}
    \label{fig: figrdisterr}
\end{figure}

We varied the sensor distance $r_C$ from the origin in the range from 5 cm to 10 cm. This range approximately covers sensor distances applicable to both infant to adult head sizes, for both OPMs and SQUID sensors that are placed as close to the head as possible. As shown before, shorter brain-to-sensor distances will allow us to detect higher spatial frequencies, so we expect higher relative errors similar to Figure \ref{fig: figm0err}, and this is, indeed,  verified by Figure \ref{fig: figrdisterr}.

Again, point-like sensors had highest relative error -- in this case, for $l=8$, there is a significant error of more than $40\%$ for sensors located at around 6 cm from the origin, the approximate size of an infant head. Since the truncation order $L=8$ is for average adult heads and lower than what is required for infant heads, this suggests that the flux basis with appropriate $L$ truncation for infants have even higher errors for higher frequency components, as indicated by our results from Figure \ref{fig: figm0err}. This may compromise spatial resolution of the source estimates in infant MEG due to the small head size. As before, for $l=8$, the 9-point cubature for square sensors and 7- and 21-point cubatures for circular sensors had small errors $<2\%$.

\subsection{Increased sensor size} \label{results3}
\begin{figure}[ht]
    \centering
    \begin{subfigure}[t]{0.49\textwidth}
        \centering
         \includegraphics[width=\textwidth]{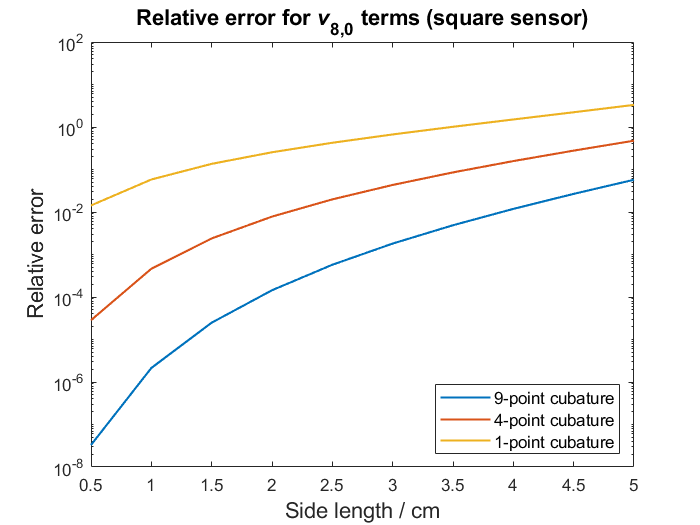}
         \caption{}
    \end{subfigure}
    \begin{subfigure}[t]{0.49\textwidth}
        \centering             
        \includegraphics[width=\textwidth]{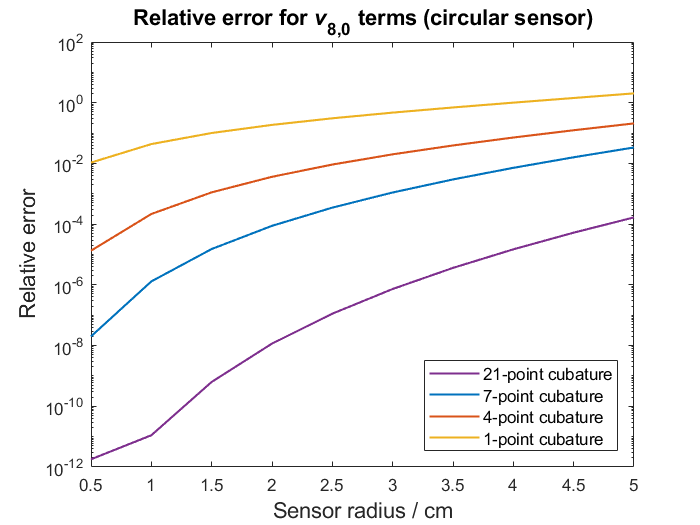}
        \caption{}
    \end{subfigure}
    \caption{Relative error plots for tangential (a) square sensors and (b) circular sensors along the $z$-axis. The errors increase as $l$ increases, which means that flux calculations for higher frequencies become more inaccurate. Fewer sampling points result in larger errors as well.}
    \label{fig: figsidelengtherr}
\end{figure}

We also considered varying sensor sizes for square half-width and circle radius sizes $d = 0.25$ cm to 2.5 cm. With increasing sensor size, the cubature sampling points get farther apart from each other and hence are not able to represent higher spatial frequency components well. We thus expect higher relative errors, which Figure \ref{fig: figsidelengtherr} verifies.

For recent OPM forward calculations and recently-built OPMs, the cylindrical cap surface has half-length to be between 5 mm to 1 cm \cite{opm1,opm4}, which is smaller than that of typical SQUID pick-up loops. The lower relative errors for smaller loops suggest that this may mitigate higher errors due to higher sensitivities as discussed in the previous two error considerations.

\subsection{Basis subspace angles between cubatures}

For this subsection only, we used the standard placement of 102 square magnetometers for the Elekta Neuromag TRIUX system. Each sensor has half-width $d=1$ cm as before. We then calculated the subspace angle for each $l$ degree (all $m$'s included), between the 1-, 4- and 9-point cubatures relative to the 21-point cubature.

\begin{figure}[ht]
    \centering
    \includegraphics[width=0.5\textwidth]{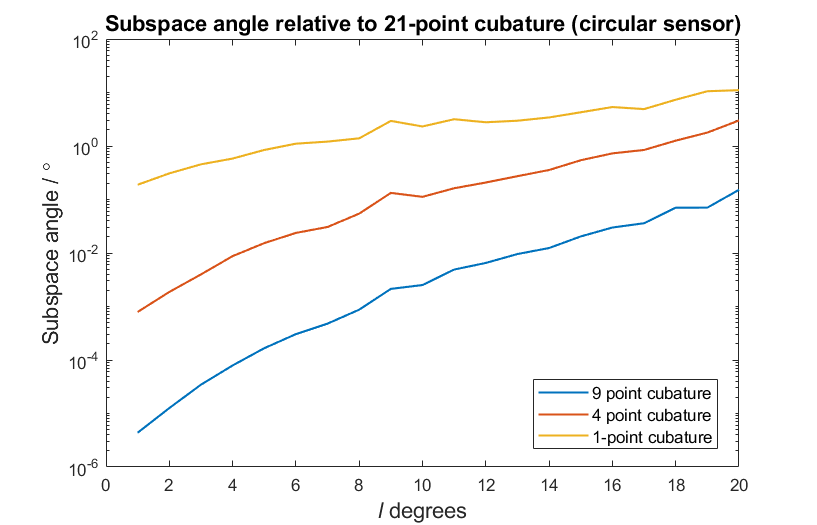}
    \caption{Subspace angle for each $l$ degree portion of the signal basis (including all $m$'s) for circular sensors, relative to the 21-point cubature basis.}
    \label{fig: cubatureangles}
\end{figure}

The impact of surface integral discretization errors can be indirectly investigated with the help of the SSS basis. Since all brain signals can be satisfactorily represented with the internal SSS basis, provided that the truncation order is high enough, one can investigate errors caused by insufficient surface discretization with the help of the subspace angle of individual SSS basis vectors. In other words, the subspace angle between vectors calculated using a coarse cubature function compared to the 21-point cubature, which we have found sufficiently accurate, is a measure of signal deviation caused by discretization, calculated for the whole sensor array. Figure \ref{fig: cubatureangles} shows the subspace angles for different orders of $l$ and different number of discretization points. As expected, the angle increases with increasing $l$. 

\section{Discussion} \label{discussion}

In this paper, we have written a simplified formula for the magnetic flux basis over arbitrarily-shaped MEG sensor loops in terms of a line integral, equation \eqref{parametricvlm}. It requires finding a parametrization of the loop, and since it depends on one integration parameter instead of two, it is more easily evaluated to arbitrary accuracy by computer software. For simple geometries like a square sensor, the integral is easily computed.

We also offer an alternative way of calculating the flux for tangential circular sensors along the $z$-axis in terms of a recursion in Section \ref{circularsensoryr} that is computationally efficient and analytic -- this can be generalized to arbitrary tangential positions via a passive rotation of the $\mathbf{B}$ field, as discussed in \ref{appendixrotation} and \ref{appendixrotationcircle}. We may also apply this concept to evaluate arbitrary tangential square sensors by passively rotating the vector potential $\mathbf{A}$ to align a sensor along the $z$-axis, and then carry out the line integral. Evaluations along the $z$-axis allow certain $m$ orders of $v_{lm}$ to be ignored since they evaluate to zero due to symmetry. For tangential square sensors, flux contributions are possibly nonzero whenever $m$ is a multiple of 4 (including $m=0$), whereas for tangential circular sensors, flux is possibly nonzero only at the $m=0$ terms.

These exact evaluations (near-exact for square sensors) were then used to calculate the relative errors of currently-employed cubatures that approximate the magnetic flux basis. The results show that the relative error increases for increased spatial frequency components. A higher number of cubature points decrease these errors, thus we suggest whenever possible to either use high numbers of cubature points, or calculate the integrals exactly. Decreased sensor distance also increases errors, which makes sense intuitively since higher frequencies can be measured. Increased sensor size increases errors due to the cubature points being sparser and hence unable to resolve higher frequency components well. For especially high frequencies, we recommend strictly using exact calculations, since we may experience signal aliasing when evaluating with the finite cubature sampling points.

Our results also show that the 9-point cubature for square sensors and 7-point cubature for circular sensors seem to provide sufficiently accurate evaluations of the signal for current generic SQUID sensor configurations; however, in the future when sensors will be more sensitive and can be placed closer to the head, they may not be sufficiently accurate. For example, forward calculations of signals for on-scalp sensors on infant heads may suffer from very large errors. 

We note that our set-ups and standards for accuracy do not necessarily apply for all purposes; as such, we have provided our Matlab codes as supplementary material so that interested readers may modify it accordingly to their needs. The codes follow the procedure as in Sections \ref{results1}-\ref{results3} and reproduce Figures \ref{fig: figm0err}-\ref{fig: cubatureangles} and Figure \ref{fig: meven}.

As additional consideration, we have presented a possible starting point for flux calculations in cylindrical harmonics in \ref{appendixcylinharm}. Since OPM-based systems typically utilize cylindrical sensing volumes, cylindrical harmonics may offer a more natural description of the signals than spherical harmonics.

\section{Conclusion} \label{conclusion}

In this paper, we have provided a review of flux calculations using the VSH expansion formalism. We have also simplified the surface flux integral of pick-up loops into a line integral as well as a recursion (latter for tangential circular sensors only). The results of different cubature approximation accuracies when varying certain sensor configuration properties were then presented. Our spatial frequency specific formalism is expected to be especially informative for novel sensor arrays that provide information on finer spatial fine details than conventional systems.

\appendix 

\section{Vector spherical harmonics} \label{appendixVSH}
The vector spherical harmonics as defined in \cite{ELH} are
\begin{align}
    \bm{\nu}_{lm} (\theta,\phi) &= -(l+1) Y_{lm} (\theta,\phi) \mathbf{e}_r + \frac{\partial Y_{lm} (\theta,\phi)}{\partial \theta} \mathbf{e}_\theta + \frac{i m Y_{lm} (\theta,\phi)}{\sin \theta} \mathbf{e}_\phi  \nonumber \\
    &\equiv \sqrt{(l+1)(2l+1)} \mathbf{V}_{lm} (\theta,\phi) \\
    \bm{\omega}_{lm} (\theta,\phi) &= l Y_{lm} (\theta,\phi) \mathbf{e}_r + \frac{\partial Y_{lm} (\theta,\phi)}{\partial \theta} \mathbf{e}_\theta + \frac{i m Y_{lm} (\theta,\phi)}{\sin \theta} \mathbf{e}_\phi \nonumber \\
    &\equiv \sqrt{l(2l+1)} \mathbf{W}_{lm} (\theta,\phi) \\
    \mathbf{x}_{lm}(\theta,\phi) &= \frac{m}{\sin \theta} Y_{lm} \mathbf{e}_\theta + i \frac{\partial Y_{lm}}{\partial \theta} \mathbf{e}_\phi  \nonumber \\
    &\equiv -\sqrt{l(l+1)} \mathbf{X}_{lm} (\theta,\phi).
\end{align}

\section{Passive rotations of vector fields} \label{appendixrotation}
The rotation matrix $\mathcal{R}$ for a proper rotation about $\mathbf{k} = (k_x, k_y, k_Z)$ with $\abs{\mathbf{k}} = k = 1$ by an angle $\gamma$ can be written using Rodrigues' rotation formula,
\begin{equation}
    \mathcal{R} = \begin{bmatrix}
    \cos \gamma + k_x^2 (1-\cos \gamma) & k_x k_y (1-\cos \gamma) - k_z \sin \gamma & k_x k_z (1-\cos \gamma) + k_y \sin \gamma \\
    k_y k_x (1-\cos \gamma) + k_z \sin \gamma & \cos \gamma + k_y^2 (1-\cos \gamma) & k_y k_z (1-\cos \gamma) - k_x \sin \gamma \\
    k_z k_x (1-\cos \gamma) - k_y \sin \gamma & k_z k_y (1-\cos \gamma) + k_x \sin \gamma &  \cos \gamma + k_z^2 (1-\cos \gamma).
    \end{bmatrix}
\end{equation}
In our case, we want to passively rotate $\mathbf{B}$ such that the $z$ axis coincides with a sensor of our choice (so we use $\mathcal{R}^{-1}$ instead of $\mathcal{R}$). Let this sensor be located at $\mathbf{r}_s = (R_s, \theta_s, \phi_s)$ (Cartesian coordinates $(x_s, y_s, z_s)$). We obtain $\mathbf{k}$ via the normalized cross product between $\mathbf{r}_s$ and the the $z$ unit vector; note that $\mathbf{k}$ always lies on the $xy$ plane, so $k_z = 0$. The rotation matrix simplifies to become
\begin{equation}
    \mathcal{R} = \begin{bmatrix}
    \cos \theta_s + k_x^2 (1-\cos \theta_s) & k_x k_y (1-\cos \theta_s) & k_y \sin \theta_s \\
    k_y k_x (1-\cos \theta_s) & \cos \theta_s + k_y^2 (1-\cos \theta_s) & -k_x \sin \theta_s \\
    -k_y \sin \theta_s & k_x \sin \theta_s &  \cos \theta_s
    \end{bmatrix}.
\end{equation}
Then, the passively rotated $\mathbf{B}$ field is given by $\mathbf{B}'(\mathbf{r}') = \mathcal{R}^{-1} \mathbf{B}(\mathcal{R}\mathbf{r})$.

\section{Passive rotation of \texorpdfstring{$v_{lm}$}{} terms for tangential circular sensors along $z$-axis} \label{appendixrotationcircle}

Preserve the notations in \ref{appendixrotation}. If we passively rotate our vector fields to the new coordinates $\mathcal{R}\mathbf{r} = (R',\theta',\phi')$ about the origin so that the new $z$-axis goes through our circular sensor $\mathcal{C}'$, then the $v_{lm}$ in this case is
    \begin{align}
    v_{l,m} = \int_{\mathcal{C}'} \frac{1}{R^{l+2}}
    &\bigg[
    \mathcal{T}_{31}\left(B_1\sin\theta'\cos\phi' + B_2\cos\theta'\cos\phi' - B_3\sin\theta'\right) \nonumber
    \\
    &\, +
    \mathcal{T}_{32}\left(B_1\sin\theta'\sin\phi' + B_2\cos\theta'\sin\phi' + B_3\cos\phi'\right) ) \nonumber
    \\
    &\, +
    \mathcal{T}_{33}\left(B_1\cos\theta' - B_2\sin\theta'\right)
    \bigg]
    dS',
    \end{align}
where $\mathcal{T}_{ij}$ is the $ij^{th}$ entry of $\mathcal{R}^{-1}$, and
    \begin{equation}
    B_1 = -(l+1)Y_{lm}(\theta',\phi'),\qquad
    B_2 = \frac{\partial Y_{lm} (\theta',\phi')}{\partial \theta}
        =
        \frac{\partial Y_{lm} (\theta',\phi')}{\partial \theta'} \frac{d \theta'}{d\theta},\qquad
    B_3 = \frac{im Y_{lm}(\theta',\phi')}{\sin\theta'}.
    \end{equation}
If we further let $\mathcal{R}_{ij}$ be the $ij^{th}$ entry of $\mathcal{R}$, then
    $$
    \frac{d \theta'}{d\theta} = \mathcal{R}_{22}.
    $$
The integral above can be simplified by observing that the terms involving $\phi'$ vanish due to symmetry of the circle, implying
    \begin{equation}
    v_{l,m} = \int_{\mathcal{C}} \frac{\mathcal{T}_{33}}{R^{l+2}}
    \left[
    -(l+1)Y_{lm}(\theta',\phi')\cos\theta' - \mathcal{R}_{22}
    \frac{\partial Y_{lm} (\theta',\phi')}{\partial \theta'}
    \sin\theta'
    \right]
    dS.
    \end{equation}
Again, by the symmetry of the circle, this integral is nonzero only when $m=0$. Recalling the notation $\gamma_{s,t,u,v}$ in Section \ref{circularsensoryr}, the passively rotated $v_{l0}$ transforms into
	\begin{equation}
	v_{l,0} = 2\pi\sqrt{\frac{2l+1}{4\pi}}\mathcal{T}_{33}
	\left(
		-(l+1)r_C \gamma_{1,l+3,0,l}
		+ \mathcal{R}_{22}\gamma_{3,l+4,1,l}
	\right) .
	\end{equation}
The same recurrences established in Section \ref{circularsensoryr} can now be used to compute $v_{l,m}$ in this case.

\section{Nonzero \texorpdfstring{$v_{lm}$}{} terms of square sensors along \texorpdfstring{$z$}{}-axis} \label{appendixmzero}

We will now show that  for tangential square sensors centered along the $z$-axis, $v_{lm} = 0$ in the following two cases:
    \begin{itemize}
    \item $m$ is odd;
    \item $m=2k$, where $k$ is an odd integer satisfying $0<|2k|\leq l$.
    \end{itemize}
In other words, $v_{lm}$ may be nonzero only when $m$ is a multiple of $4$ (with $0\leq |m|\leq l$). Intuitively, this is due to the dihedral $D_4$-symmetry of the square, and we provide mathematical justification for the assertion as follows.

Let us assume a similar set-up for a square sensor of half-width $d$ as in Section \ref{circularsensoryr}. Then equation \eqref{keyeqn} for $v_{lm}$ still holds, except with $\mathcal{C}$ being the square area. We note that a clean recursive formula seems to be difficult to obtain in this case, but we may still make some useful comments about which $v_{lm}$ terms are nonzero based on symmetry, like in the circular case.

Assume $m$ is odd. Pick any point $(x,y)$ in the first quadrant of the square, and compare the integrand of \eqref{keyeqn} at the four points $(x,y),(x,-y),(-x,-y),(-x,y)$. The only difference in the integrand when evaluated at any point $P$ among these four points is the exponential term $e_P = e^{im\phi_{P}}$ in $Y_{lm}$, where $\phi_P$ calculated using Equation \eqref{cartosph}. Since $\cos(m\pi) = (-1)^m = -1$, observe that
    \begin{align*}
    e_{(x,y)} &= \cos(m\phi_{(x,y)}) + i\sin(m\phi_{(x,y)}); \\   
    e_{(x,-y)} &= \cos(m\phi_{(x,y)}) - i\sin(m\phi_{(x,y)}); \\   
    e_{(-x,-y)} &= -\cos(m\phi_{(x,y)}) - i\sin(m\phi_{(x,y)}); \\   
    e_{(-x,y)} &= -\cos(m\phi_{(x,y)}) + i\sin(m\phi_{(x,y)}).
    \end{align*}
Hence the sum of these four integrands equals zero, and our conclusion follows by symmetry of the square.

Now assume $m=2k$,  where $k$ is an odd integer satisfying $0<|2k|\leq l$. As before, pick any point $Q_1=(x,y)$ in the first quadrant of the square. Now consider the points $Q_2,Q_3,Q_4$ by rotating $Q_1$ counterclockwise about the origin by an angle of $\frac{\pi}{2},\pi,\frac{3\pi}{2}$ respectively. Notice once again that the only difference in the integrand of \eqref{keyeqn} is the exponential term, and in this case
    $$
    e_{Q_1} = -e_{Q_2} = e_{Q_3} = -e_{Q_4},
    $$
implying the sum of the four integrands equals zero again.

\begin{figure}[h]
    \centering
    \begin{subfigure}[t]{0.49\textwidth}
        \centering
         \includegraphics[width=\textwidth]{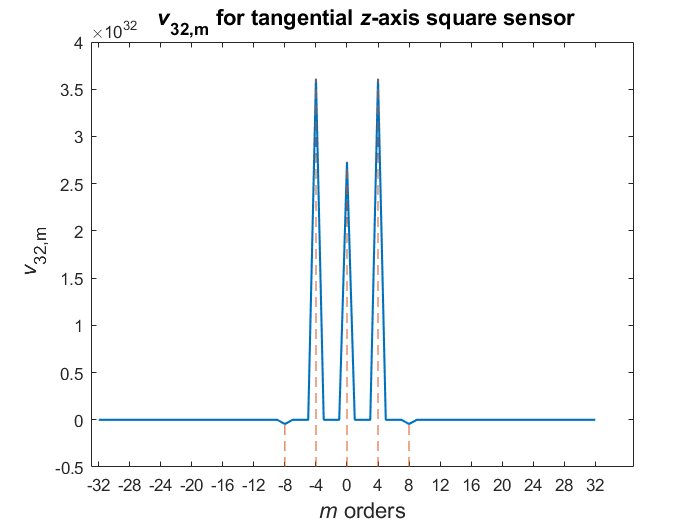}
         \caption{}
         \label{fig: mevensq}
    \end{subfigure}
    \begin{subfigure}[t]{0.49\textwidth}
        \centering             
        \includegraphics[width=\textwidth]{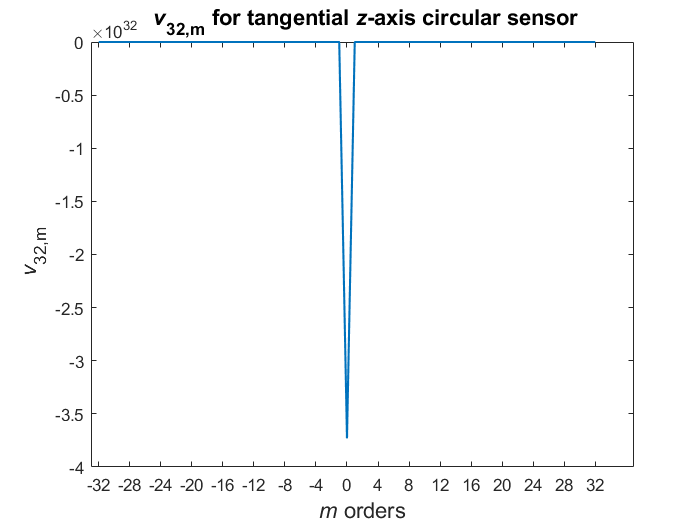}
        \caption{}
        \label{fig: mevencirc}
    \end{subfigure}
    \caption{Comparison of $v_{32,m}$ terms between tangential (a) square sensors and (b) circular sensors along the $z$-axis. The red dashed lines in (a) show that indeed, $v_{32,m} \neq 0$ only when $m$ is a multiple of 4, including $m=0$ for square sensors. (b) illustrates how only $v_{32,0}$ will be nonzero for circular sensors.}
    \label{fig: meven}
\end{figure}

The above arguments do not apply when $m$ is a multiple of 4 (including $m=0$). Indeed, in contrast to the circular case, $v_{lm}$ is not necessarily zero in this case. Figure \ref{fig: meven} is a comparison of the square and circular $v_{32,m}$ terms that illustrates this. Note that due to computer software round-off errors, the $m$ odd values may have nonzero values; however, they are insignificant compared to the $m$ even values. 

\section{Cylindrical harmonics} \label{appendixcylinharm}
We state the following result from \cite{sph2cyl1, sph2cyl2} that allows us to convert the (solid) spherical harmonics to cylindrical harmonics,
\begin{align}
    \frac{Y_{lm} (\theta,\phi)}{R^{l+1}} &= \frac{c_{lm}}{(l-m)!} e^{im\phi}\int_0^{\infty} \lambda^l e^{-\lambda z} J_m(\lambda \rho) d\lambda \label{sphtocylind1} \\
    R^lY_{lm} (\theta,\phi) &= \frac{c_{lm}(l-m)!}{2 \pi i} e^{im\phi} \int_0^{\infty} \frac{e^{\lambda z}}{\lambda^{l+1}} J_m(\lambda \rho ) d\lambda \label{sphtocylind2}
\end{align}
where $(\rho, \phi, z) = (R \sin \theta,\phi, R \cos \theta)$ are the cylindrical coordinates, $c_{lm}$ are the spherical harmonic coefficients as seen in \eqref{sphericalharmonics}, and $J_m$ are the Bessel functions of the first kind. Denoting $K_{lm}$ and $L_{lm}$ as the definite integrals in \eqref{sphtocylind1} and \eqref{sphtocylind2}, respectively,  the non-normalized (solid) cylindrical harmonics are defined as $e^{im\phi} K_{lm}(\rho,z)$ and $e^{im\phi} L_{lm}(\rho,z)$. One can also arrive at these expressions by solving Laplace's equation in cylindrical coordinates from the beginning -- see for example \cite{jackson}. 

The scalar potential in cylindrical harmonics is thus
\begin{equation}
    V (\mathbf{r}) = \sum_{l=1}^\infty \sum_{m=-l}^l \alpha_{lm} \frac{c_{lm} }{(l-m)!} e^{im\phi} K_{lm}(\rho,z) + \sum_{l=1}^\infty \sum_{m=-l}^l \beta_{lm} \frac{c_{lm}(l-m)! }{2 \pi i} e^{im\phi} L_{lm}(\rho,z).
\end{equation}
We observe that
\begin{align}
    \nabla \left( e^{im\phi}K_{lm}\right) = e^{im\phi} \left( \frac{\partial K_{lm}}{\partial \rho}  \hat{\bm{\rho}} + \frac{im}{\rho} K_{lm} \hat{\bm{\phi}} - K_{l+1,m} \hat{\mathbf{z}} \right) \equiv \mathbf{K}_{l+1,m}\\
    \nabla \left( e^{im\phi}L_{lm}\right) = e^{im\phi} \left(\frac{\partial L_{lm}}{\partial \rho}  \hat{\bm{\rho}} + \frac{im}{\rho} L_{lm} \hat{\bm{\phi}} + L_{l-1,m} \hat{\mathbf{z}}\right) \equiv \mathbf{L}_{l-1,m},
\end{align}
where we have defined $\mathbf{K}_{lm}$ and $\mathbf{L}_{lm}$ as the vector cylindrical harmonics. We may evaluate the derivative with respect to $\rho$ using
\begin{equation}
    \frac{d}{d\rho} J_m (\lambda \rho) =  \frac{d}{d (\lambda\rho)} \frac{d(\lambda\rho)}{d\rho} J_m (\lambda\rho) = \frac{\lambda}{2} \left(J_{m-1} (\lambda \rho) - J_{m+1}(\lambda\rho)\right),
\end{equation}
which gives 
\begin{align}
    \frac{\partial K_{lm}}{\partial \rho} &= \frac{1}{2} \left[K_{l+1,m-1} - K_{l+1,m+1}\right] \\
    \frac{\partial L_{lm}}{\partial \rho} &= \frac{1}{2} \left[L_{l-1,m-1} - L_{l-1,m+1}\right].
\end{align}
Then, the magnetic field is 
\begin{equation}
    \mathbf{B} (\mathbf{r}) = - \mu_0 \sum_{l=1}^\infty \sum_{m=-l}^l \alpha_{lm} \frac{c_{lm}}{(l-m)!} \mathbf{K}_{l+1,m}(\rho,\phi,z) - \mu_0 \sum_{l=1}^\infty \sum_{m=-l}^l \beta_{lm} \frac{c_{lm}(l-m)! }{2 \pi i} \mathbf{L}_{l-1,m}(\rho,\phi,z).
\end{equation}
If we assume a tangential sensor, then we may passively rotate this magnetic field (\ref{appendixrotation}) so that the $z$-axis aligns with the sensor, similar to Section \eqref{circularsensoryr}. For simplicity let us assume the sensor is already aligned along the $z$-axis. In this case, the volume integration bounds are easily specified. 

As a comparison with the surface flux equation \eqref{phiinout}, we reduce the volume integral to a surface integral by integrating with respect to the cylinder depth $z$, from the inner cylinder cap height $z_1$ to the outer cap height $z_2$,
\begin{align}
    \bm{\Phi}_{surf} \left(\mathbf{r}\right) = &\Bigg[ \mu_0 \sum_{l=1}^\infty \sum_{m=-l}^l \alpha_{lm} \int_{\mathcal{C}} \frac{c_{lm}}{(l-m)!} \mathbf{K}_{lm} (\rho, \phi, z) \cdot \mathbf{n}' dS \nonumber \\
    &- \mu_0 \sum_{l=1}^\infty \sum_{m=-l}^l \beta_{lm} \int_{\mathcal{C}} \frac{c_{lm}(l-m)! }{2 \pi i} \mathbf{L}_{lm} (\rho, \phi, z) \cdot \mathbf{n}' dS \Bigg]\Bigg|_{z=z_1}^{z_2}.
\end{align}
In this form, if one were to use SSS with cylindrical harmonics, the basis matrices would be specified by the integrals. In principle, the bounds for the integrals for the cylinder cap surfaces may be specified in a straightforward manner. Evaluating this integral in a simpler way is a topic for future study.

\section*{Acknowledgements}

S. Taulu was supported by the Bezos Family Foundation and the R.\ B.\ and Ruth H.\ Dunn Charitable Foundation.

\end{document}